%% file: main.tex
\documentclass[10pt,twocolumn,letterpaper]{article}

\usepackage[accsupp]{axessibility}
\usepackage[pagenumbers]{arxiv}              % To produce the CAMERA-READY version
\usepackage[dvipsnames]{xcolor}
\usepackage{wrapfig}

\definecolor{cvprblue}{rgb}{0.21,0.49,0.74}

\usepackage[pagebackref,breaklinks,colorlinks,citecolor=cvprblue]{hyperref}
\usepackage{bm}
\usepackage{booktabs}
\newcommand\blfootnote[1]{%
\begingroup
\renewcommand\thefootnote{}\footnote{#1}%
\addtocounter{footnote}{-1}%
\endgroup
}

\def\paperID{4555} % *** Enter the Paper ID here
\def\confName{CVPR}
\def\confYear{2024}

\newcommand{\METHOD}{PhysGaussian}
\title{\METHOD: Physics-Integrated 3D Gaussians for Generative Dynamics}

\author{Tianyi Xie$^{1*}$  \quad Zeshun Zong$^{1*}$\quad Yuxing Qiu$^{1*}$ \quad Xuan Li$^{1*}$ \\
Yutao Feng$^{2,3}$ \quad Yin Yang$^{3}$ \quad Chenfanfu Jiang$^{1}$ \\
$^{1}$ UCLA, $^{2}$ Zhejiang University, $^{3}$ University of Utah
}
\begin{document}
% \maketitle

\twocolumn[{%
\renewcommand\twocolumn[1][]{#1}%
\maketitle
\begin{center}
    \centering
    \captionsetup{type=figure}
    \includegraphics[width=\textwidth, trim=0 0 0 0, clip]{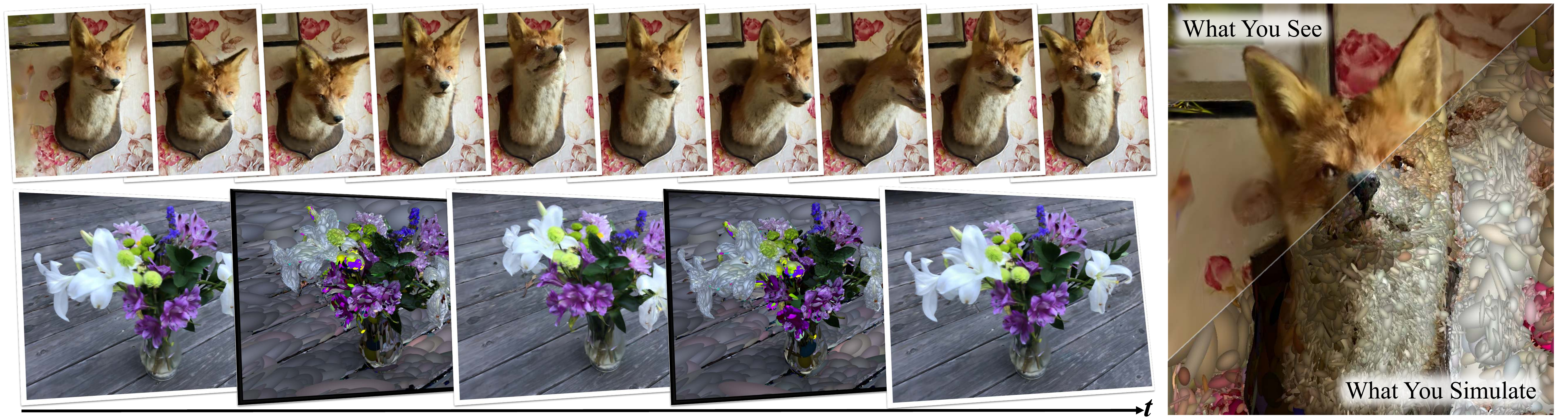}
    \captionof{figure}{\textbf{PhysGaussian} is 
    % an efficient simulation and rendering pipeline with a unified 3D Gaussian representation.}
        a unified simulation-rendering pipeline based on  3D Gaussians  and continuum mechanics.}
    %are seamlessly simulatable and renderable with novel motion and views.}
\label{fig:teaser}
\end{center}
}]
% \begin{figure*}
%     \centering
%     \includegraphics[width=\linewidth]{figures/teaser.pdf}
%     \caption{\textbf{PhysGaussian.} 
%     \label{fig:teaser}}
% \end{figure*}

\begin{abstract}
%\vspace{-10pt}
\blfootnote{* indicates equal contributions.} 
We introduce \METHOD, a new method that seamlessly integrates physically grounded Newtonian dynamics within 3D Gaussians to achieve high-quality novel motion synthesis. Employing a custom Material Point Method (MPM), our approach enriches 3D Gaussian kernels with physically meaningful kinematic deformation and mechanical stress attributes, all evolved in line with continuum mechanics principles. A defining characteristic of our method is the seamless integration between physical simulation and visual rendering: both components utilize the same 3D Gaussian kernels as their discrete representations. This negates the necessity for triangle/tetrahedron meshing, marching cubes, ``cage meshes,'' or any other geometry embedding, highlighting the principle of ``what you see is what you simulate (WS$^2$).'' 
Our method demonstrates exceptional versatility across a wide variety of materials--including elastic entities, plastic metals, non-Newtonian fluids, and granular materials--showcasing its strong capabilities in creating diverse visual content with novel viewpoints and movements. Our project page is at: \url{https://xpandora.github.io/PhysGaussian/}.
\end{abstract}

% Illustrating its robustness, 
%Our approach excels across a spectrum of materials, spanning elastic objects, plastic metals, non-Newtonian fluids, to granular substances.
%This accentuates its potential to harmonize physics-based simulations with visual renderings for AI-generated visual content.
\vspace{-8pt}
\section{Introduction}

Recent strides in Neural Radiance Fields (NeRFs) have showcased significant advancements in 3D graphics and vision \cite{mildenhall2021nerf}. Such gains have been further augmented by the cutting-edge 3D Gaussian Splatting (GS) framework \cite{kerbl20233d}. Despite many achievements, a noticeable gap remains in the application towards generating novel dynamics. While there exist endeavors that generate new poses for NeRFs, they typically cater to quasi-static geometry shape editing tasks and often require meshing or embedding visual geometry in coarse proxy meshes such as tetrahedra \cite{yuan2022nerf,xu2022deforming,peng2022cagenerf,NeRFshop23}.

Meanwhile, the traditional physics-based visual content generation pipeline has been a tedious multi-stage process: constructing the geometry, making it simulation-ready (often through techniques like tetrahedralization), simulating it with physics, and finally rendering the scene. This sequence, while effective, introduces intermediary stages that can lead to discrepancies between simulation and final visualization. Even within the NeRF paradigm, a similar trend is observed, as the rendering geometry is embedded into a simulation geometry. This division, in essence, contrasts with the natural world, where the physical behavior and visual appearance of materials are intrinsically intertwined. 
Our overarching philosophy seeks to align these two facets by advocating for a unified representation of a material substance, employed for both simulation and rendering. In essence, our approach champions the principle of \textit{``what you see is what you simulate'' (WS$^2$)} \cite{muller2016simulating}, aiming for a more coherent integration of simulation, capturing, and rendering.

Building towards this goal, we introduce \METHOD: physics-integrated 3D Gaussians for generative dynamics. This novel approach empowers 3D Gaussians to encapsulate physically sound Newtonian dynamics, including realistic behaviors and inertia effects inherent in solid materials. More specifically, we impart physics to 3D Gaussian kernels, endowing them with kinematic attributes such as velocity and strain, along with mechanical properties like elastic energy, stress, and plasticity. Notably, through continuum mechanics principles and a custom Material Point Method (MPM), \METHOD \ ensures that both physical simulation and visual rendering are driven by 3D Gaussians. This eradicates the necessity for any embedding mechanisms, thus eliminating any disparity or resolution mismatch between the simulated and the rendered. 

We present \METHOD's versatile adeptness in synthesizing generative dynamics across various materials, such as elastic objects, metals, non-Newtonian viscoplastic substances (e.g. foam or gel), and granular mediums (e.g. sand or soil). To summarize, our contributions include
\begin{itemize}
\item \textbf{Continuum Mechanics for 3D Gaussian Kinematics:} 
% We present a methodology for evolving 3D Gaussian kernels and their associated spherical harmonics in a manner that ensures their deformation responses are appropriate and accurate within the context of continuum mechanics.
%We introduce a continuum mechanics-based strategy tailored for 3D Gaussian kernels and their associated spherical harmonics. \zzs{the following sentence is weird, respond to what? what does appropriate mean?} Our approach governs their evolution, ensuring they respond appropriately in the simulated deformation field.
We introduce a continuum mechanics-based strategy tailored for evolving 3D Gaussian kernels and their associated spherical harmonics in physical Partial Differential Equation (PDE)-driven displacement fields.

\item \textbf{Unified Simulation-Rendering Pipeline:} 
We present an efficient simulation and rendering pipeline with a unified 3D Gaussian representation. % The motion generation process is thus simplified by eliminating the extra effort for explicit object meshing.
Eliminating the extra effort for explicit object meshing, the motion generation process is significantly simplified. 
% Our approach introduces an efficient pipeline using 3D Gaussians for both simulation and rendering, thus streamlining the process of motion graphics generation and eliminating the need for explicitly meshing the simulated objects.

\item \textbf{Versatile Benchmarking and Experiments:} We conduct a comprehensive suite of benchmarks and experiments across various materials. Enhanced by real-time GS rendering and efficient MPM simulations, we achieve \emph{real-time} performance for scenes with simple dynamics.
\end{itemize}

\section{Related Work}
% \paragraph{Static NeRF}
\paragraph{Radiance Fields Rendering for View Synthesis.}
Radiance field methods have gained considerable interest in recent years due to their extraordinary ability to generate novel-view scenes and their great potential in 3D reconstruction. 
The adoption of deep learning techniques has led to the prominence of neural rendering and point-based rendering methods, both of which have inspired a multitude of subsequent works.
On the one hand, the NeRF framework employs a fully-connected network to model one scene \cite{mildenhall2021nerf}. The network takes spatial position and viewing direction as inputs and produces the volume density and radiance color. These outputs are subsequently utilized in image generation through volume rendering techniques. 
Building upon the achievements of NeRF, further studies have focused on enhancing reconstruction quality and improving training speeds \cite{fridovich2022plenoxels, muller2022instant, sun2022direct, barron2022mip, xu2022point, lin2022neurmips}. 
On the other hand, researchers have also investigated differentiable point-based methods for real-time rendering of unbounded scenes.
Among the current investigations, the state-of-the-art results are achieved by the recently published 3D Gaussian Splatting framework \cite{kerbl20233d}. 
Contrary to prior implicit neural representations, GS employs an explicit and unstructured representation of one scene, offering the advantage of straightforward extension to post-manipulation. 
Moreover, its fast visibility-aware rendering algorithm also enables real-world dynamics generations.

\paragraph{Dynamic Neural Radiance Field.} 
An inherent evolution of the NeRF framework entails the integration of a temporal dimension to facilitate the representation of dynamic scenes. 
For example, both \citet{pumarola2021d} and \citet{park2021nerfies} decompose time-dependent neural fields into an inverse displacement field and canonical time-invariant neural fields.
In this context, the trajectory of query rays is altered by the inverse displacement field and then positioned within the canonical space.
Subsequent studies have adhered to the aforementioned design when exploring applications related to NeRF deformations, such as static scene editing and dynamic scene reconstruction \cite{li2023climatenerf, peng2022cagenerf, yuan2022nerf, chen2022virtual, qiao2022neuphysics, qiao2023dynamic, liu2023neural}. 
Additionally, \citet{yuan2022nerf, qiao2022neuphysics, liu2023neural} have contributed to the incorporation of physics-based deformations into the NeRF framework.
However, the effectiveness of these methodologies relies on the usage of exported meshes derived from NeRFs.  
To circumvent this restriction, explicit geometric representations have been explored for forward displacement modeling \cite{xu2022point, kerbl20233d}.
In particular, \citet{chen2023neuraleditor, luiten2023dynamic, yang2023deformable, wu20234d, yang2023real} directly manipulate NeRF fields. \citet{li2023pacnerf} extends this approach by including physical simulators to achieve more dynamic behaviors. 
In this study, we leverage the explicit 3D Gaussian Splatting ellipsoids as a unified representation for both physics and graphics.
In contrast to previous dynamic GS frameworks, which either maintain the shapes of Gaussian kernels or learn to modify them, our approach uniquely leverages the first-order information from the displacement map (deformation gradient) to assist dynamic simulations. In this way, we are able to deform the Gaussian kernels and seamlessly integrate the simulation within the GS framework.

\begin{figure*}[t]
    \centering
    \includegraphics[width=1\linewidth]{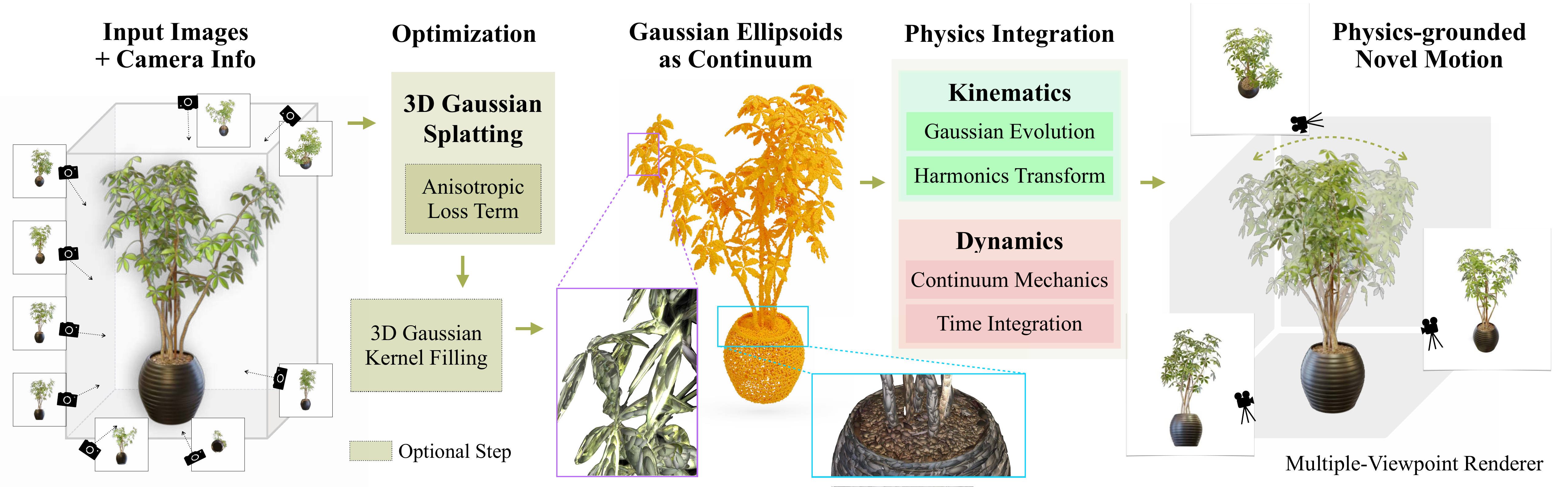}
    \caption{\textbf{Method Overview.} \METHOD~ is a unified simulation-rendering pipeline that incorporates 3D Gaussian splatting representation and continuum mechanics to generate physics-based dynamics and photo-realistic renderings simultaneously and seamlessly.}
    \label{fig:overview}
\end{figure*}

\paragraph{Material Point Method.} The Material Point Method (MPM) is a widely used simulation framework for a broad range of multi-physics phenomena \cite{hu2018moving}. The inherent capability of the MPM system allows for topology changes and frictional interactions, making it suitable for simulating various materials, including but not limited to elastic objects, fluids, sand, and snow \cite{stomakhin2013material, jiang2015affine, klar2016drucker}. MPM can also be expanded to simulate objects that possess codimensional characteristics \cite{jiang2017anisotropic}. In addition, the efficacy of utilizing GPU(s) to accelerate MPM implementations has also been demonstrated in \cite{gao2018gpu, hu2019taichi, wang2020massively, qiu2023sparse}. 
Owing to its well-documented advantages, we employ the MPM to support the latent physical dynamics.   
This choice allows us to efficiently import dynamics into various scenarios with a shared particle representation alongside the Gaussian Splatting framework.

\section{Method Overview}
\label{sec:overview}

We propose \METHOD\ (\cref{fig:overview}), a unified simulation-rendering framework for generative dynamics based on continuum mechanics and 3D GS. Adopted from \citet{kerbl20233d}, we first reconstruct a GS representation of a static scene, with an optional anisotropic loss term to regularize over-skinny kernels. These Gaussians are viewed as the discretization of the scene to be simulated. Under our novel kinematics, we directly splat the deformed Gaussians for photo-realistic renderings. For better physics compliance, we also optionally fill the internal regions of objects. We detail these in this section.

\subsection{3D Gaussian Splatting}
\label{sec:3d_gs}
3D Gaussian Splatting method \cite{kerbl20233d} reparameterizes NeRF \cite{mildenhall2021nerf} using a set of unstructured 3D Gaussian kernels $\{\bm x_p, \sigma_p, \bm A_p, \mathcal{C}_p\}_{p\in \mathcal{P}}$, where $\bm x_p$, $\sigma_p$, $\bm A_p$, and $\mathcal{C}_p$ represent the centers, opacities, covariance matrices, and spherical harmonic coefficients of the Gaussians, respectively. To render a view, GS projects these 3D Gaussians onto the image plane as 2D Gaussians, differing from traditional NeRF techniques that emit rays from the camera.  The final color of each pixel is computed as
\begin{equation}
    \bm C = \sum_{k\in \mathcal{P}} \alpha_k \text{SH}(\bm d_k; \mathcal{C}_k)  \prod_{j=1}^{k-1} (1-\alpha_j).
\end{equation}
Here $\alpha_k$ represents the $z$-depth ordered effective opacities, \ie, products of the 2D Gaussian weights and their overall opacities $\sigma_k$; $\bm d_k$ stands for the view direction from the camera to $\bm x_k$. Per-view optimizations are performed using $L_1$ loss and SSIM loss. This explicit representation of the scene not only significantly accelerates training and rendering speeds, but also enables direct manipulation of the NeRF scene. The data-driven dynamics are supported by  making $\bm x_p, A_p$ time-dependent \cite{wu20234d} and minimizing rendering losses over videos. In \cref{sec:3d_gs}, we show that this time-dependent evolution can be given by the continuum deformation map.

\subsection{Continuum Mechanics} \label{sec:continuum}

Continuum mechanics describes motions by a time-dependent continuous deformation map $\bm x=\bm  \phi(\bm X,t)$ between the undeformed material space $\Omega^0$ and the deformed world space $\Omega^t$ at time $t$. The deformation gradient 
$F(\bm X,t)=\nabla_{\bm X}\bm  \phi(\bm X,t)$
encodes local transformations including stretch, rotation, and shear \cite{bonet1997nonlinear}. The evolution of the deformation $\bm \phi$ is governed by the conservation of mass and momentum. 
Conservation of mass ensures that the mass within any infinitesimal region $B_{\epsilon}^0 \in \Omega^0$ remains constant over time:
\begin{equation}
    \int_{B_{\epsilon}^t} \rho(\bm x,t) \equiv \int_{B_{\epsilon}^0} \rho(\bm \phi^{-1}(\bm x,t), 0),
\end{equation}
% $\int_{B_{\epsilon}^t} \rho(\bm x,t) \equiv \int_{B_{\epsilon}^0} \rho(\bm \phi^{-1}(\bm x,t), 0) $, 
where $B_{\epsilon}^t = \bm \phi(B_{\epsilon}^0, t)$ and $\rho(\bm x,t)$ is the density field characterizing material distribution.  Denoting the velocity field with $\bm v(\bm x,t) $, the conservation of momentum is given by 
\begin{equation}
    \rho(\bm x,t) \dot{\bm v}(\bm x,t) = \nabla\cdot\bm \sigma(\bm x,t) + \bm f^{\text{ext}},
\end{equation}
where $\bm \sigma = \frac1{\operatorname{det}(\bm F)}\frac{\partial \Psi}{\partial \bm F}(\bm F^E) {\bm F^E}^T$ is the Cauchy stress tensor associated with a hyperelastic energy density $\Psi(\bm F)$, and $\bm f^{\text{ext}}$ is the external force per unit volume \cite{bonet1997nonlinear,jiang2016material}. Here the total deformation gradient can be decomposed into an elastic part and a plastic part $\bm F = \bm F^E \bm F^P$ to support permanent rest shape changes caused by plasticity. The evolution of $\bm F^E$ follows some specific plastic flow such that it is always constrained within a predefined elastic region \cite{bonet1997nonlinear}. 

% \TODO{a super dense paragraph on explicit MPM steps containing expressions. mention damping mechanisim as well.}

\subsection{Material Point Method} 
\label{sec:mpm}

Material Point Method (MPM) solves the above governing equations by combining the strengths of both Lagrangian particles and Eulerian grids \cite{stomakhin2013material, jiang2016material}. The continuum is discretized by a collection of particles,  each representing a small material region. These particles track several time-varying Lagrangian quantities such as position $\bm x_p$, velocity $\bm v_p$, and deformation gradient $\bm F_p$. 
The mass conservation in Lagrangian particles ensures the constancy of total mass during movement. Conversely, momentum conservation is more natural in Eulerian representation, which avoids mesh construction. We follow \citet{stomakhin2013material} to integrate these representations using $C^1$ continuous B-spline kernels for two-way transfer.
From time step $t^n$ to $t^{n+1}$, the momentum conservation, discretized by the forward Euler scheme, is represented as
\begin{equation}
\resizebox{.9\linewidth}{!}{
    $\frac{m_i}{\Delta t} (\bm v^{n+1}_i - \bm v^{n}_i) = -\sum_p V_p^0 \frac{\partial \Psi}{\partial \bm F}(\bm F^{E, n}_p) {\bm F^{E, n}_p}^T \nabla w_{ip}^n + \bm f^{ext}_i$.
}
\label{eq:mpm_discretization}
\end{equation}
Here $i$ and $p$ represent the fields on the Eulerian grid and the Lagrangian particles respectively; $w_{ip}^n$ is the B-spline kernel defined on $i$-th grid evaluated at $\bm x_p^n$; $V_p^0$ is the initial representing volume, and $\Delta t$ is the time step size. The updated grid velocity field $\bm v_i^{n+1}$ is transferred back onto particle to $v_p^{n+1}$, updating the particles' positions to 
    $\bm x_p^{n+1} = \bm x_p^n + \Delta t \bm v_p^{n+1}$.
We track $\bm F^E$ rather than both $\bm F$ and $\bm F^P$ \cite{simo2006computational}, which is updated by
% \begin{equation}
% \begin{aligned}
%     \bm F^{E, n+1}_p &= (\bm I + \Delta t \nabla \bm v_p)\bm F^{E, n}_p \\
%             &= (\bm{I}+\Delta t \sum_i \bm{v}_i^{n+1}{\nabla w_{i p}^n}^T) \bm F_p^{E,n}
% \end{aligned}
% \end{equation}
% \begin{equation}
%     \bm F^{E, n+1}_p = (\bm{I}+\Delta t \sum_i \bm{v}_i^{n+1}{\nabla w_{i p}^n}^T) \bm F_p^{E,n}
% \end{equation}
$\bm F^{E, n+1}_p = (\bm I + \Delta t \nabla \bm v_p)\bm F^{E, n}_p = (\bm{I}+\Delta t \sum_i \bm{v}_i^{n+1}{\nabla w_{i p}^n}^T) \bm F_p^{E,n}$ 
and regularized by an additional return mapping to support plasticity evolution:
$\bm F^{E, n+1}_p \leftarrow  \mathcal{Z}(\bm F^{E, n+1}_p)$.
Different plasticity models define different return mappings. We refer to the supplemental document for details of the simulation algorithm and different return mappings.

\subsection{Physics-Integrated 3D Gaussians}
\label{sec:physics_GS} 
We treat Gaussian kernels as discrete particle clouds for spatially discretizing the simulated continuum.
As the continuum deforms, we let the Gaussian kernels deform as well. 
However, for a Gaussian kernel defined at $\bm X_p$ in the material space, $G_p(\bm X) = e^{-\frac12 (\bm X - \bm X_p)^T \bm A^{-1}_p(\bm X - \bm X_p)}$,
the deformed kernel under the deformation map $\bm \phi (\bm X, t)$,
\begin{equation}
    G_p(\bm x, t) = e^{-\frac12 (\bm \phi^{-1}(\bm x, t) - \bm X_p)^T \bm A^{-1}_p(\bm \phi^{-1}(\bm x, t) - \bm X_p)}
\end{equation}
is not necessarily Gaussian in the world space, which violates the requirements of the splatting process. Fortunately, if we assume particles undergo local affine transformations characterized by the first-order approximation
\begin{equation}
\tilde{\phi}_p(\bm X, t)= \bm x_p + \bm F_p (\bm X - \bm X_p),
\end{equation}
the deformed kernel becomes Gaussian as desired:
\begin{equation}
     G_p(\bm x, t) = e^{-\frac12 (\bm x - \bm x_p)^T (\bm F_p\bm A_p\bm F_p^{T})^{-1}(\bm x - \bm x_p)}.
\end{equation}
\begin{wrapfigure}{r}{0.4\linewidth}
\vspace{-1em}
\hspace{-1.5em}
    \includegraphics[width=1.1\linewidth]{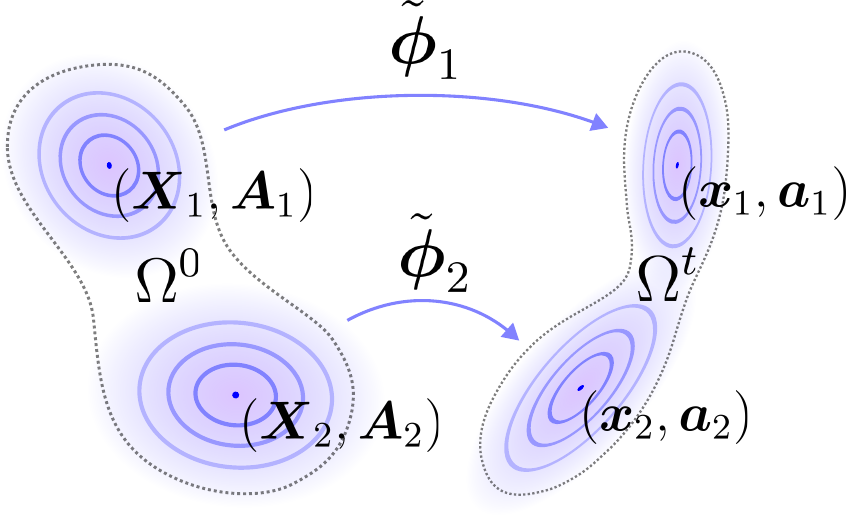}
  \vspace{-1em}
\end{wrapfigure}
This transformation naturally provides a time-dependent version of $\bm x_p$ and $\bm A_p$ for the 3D GS framework:
\begin{equation}
\small
\begin{split}
    &\bm x_p(t) = \bm \phi(\bm X_p, t), \\
    &\bm a_p(t) =  \bm F_p(t)\bm A_p\bm F_p(t)^{T}.
\end{split}
\end{equation}
In summary, given the 3D GS of a static scene $\{\bm X_p, \bm A_p, \sigma_p, \mathcal{C}_p\}$, we use simulation to dynamize the scene by evolving these Gaussians to produce dynamic Gaussians $\{\bm x_p(t), \bm a_p(t), \sigma_p, \mathcal{C}_p\}$. Here we assume that the opacity and the coefficients of spherical harmonics are invariant over time, but the harmonics will be rotated as discussed in the next section. We also initialize other physical quantities in \cref{eq:mpm_discretization}: the representing volume of each particle $V_p^0$ is initialized as background cell volume divided by the number of contained particles; the mass $m_p$ is then inferred from user-specified density $\rho_p$ as $m_p = \rho_p V_p^0$. To render these deformed Gaussian kernels, we use the splatting from the original GS framework \cite{kerbl20233d}. It should be highlighted that the integration of physics into 3D Gaussians is  \textbf{seamless}: on the one hand, the Gaussians themselves are viewed as the discretization of the continuum, which can be simulated directly; on the other hand, the deformed Gaussians can be directly rendered by the splatting procedure, avoiding the need for commercial rendering software in traditional animation pipelines. Most importantly, we can directly simulate scenes reconstructed from real data, 
achieving WS$^2$.
%highlighting our principle of ``what you see is what you simulate (WS$^2$).''

% \begin{remark}
% \TODO{Highlight that Point-based NeRF frameworks only track particle positions. They do not track first-order information (local distortion, or deformation gradient) or they simply assume rigid motions. Simulation frameworks have this information by nature.}
% \end{remark}

\subsection{Evolving Orientations of Spherical Harmonics}
\label{sec:render}
\begin{wrapfigure}{r}{0.45\linewidth}
\vspace{-0.5em}
\hspace{-1.5em}
    \includegraphics[width=1.1\linewidth]{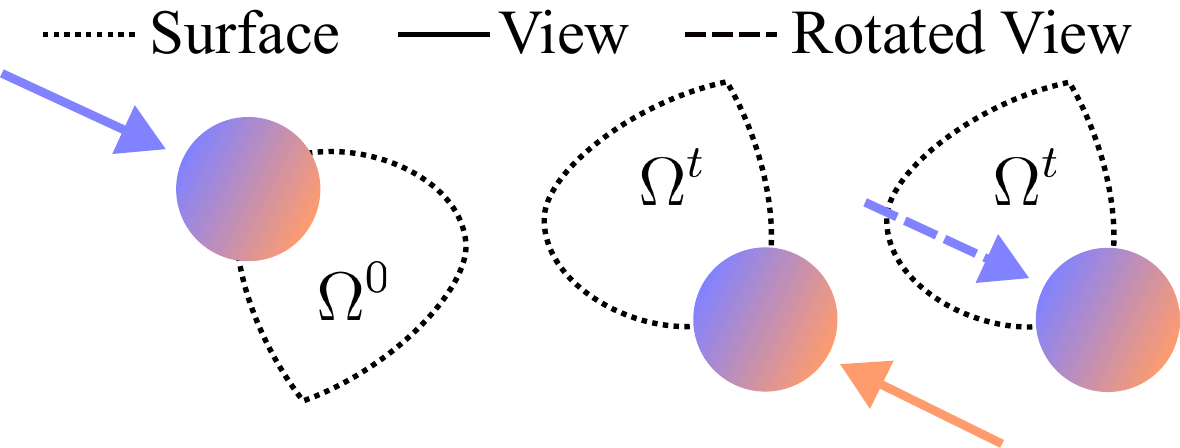}
  \vspace{-0.5em}
\end{wrapfigure}
Rendering the world-space 3D Gaussians can already obtain high-quality results. However, when the object undergoes rotations, the spherical harmonic bases are still represented in the material space, resulting in varying appearances even if the view direction is fixed relatively to the object. The solution is simple: when an ellipsoid is rotated over time, we rotate the orientations of its spherical harmonics as well. However, the bases are hard-coded inside the GS framework. We equivalently achieve this evolution by applying inverse rotation on view directions. This effect is illustrated in the inset figure. We remark that the rotation of view directions is not considered in \citet{wu20234d}. \citet{chen2023neuraleditor} tackles this issue in the Point-NeRF framework, but requires tracking of surface orientation. In our framework, the local rotation is readily obtained in the deformation gradient $\bm F_p$. 
Denote $f^0(\bm d)$ as a spherical harmonic basis in material space, with $\bm d$ being a point on the unit sphere (indicating view direction). The polar decomposition, $\bm F_p = \bm R_p\bm S_p$, leads us to the rotated harmonic basis:
\begin{equation}
    f^{t}(\bm d) = f^0(\bm R^{T}\bm d).
\end{equation}

% \TODO{This section may be optional to leave more space for results. }
% \todo{j: keep contribution sections. save space by shortening sentences, especially in mpm and continuum sections.}

\subsection{Incremental Evolution of Gaussians}
\label{sec:incremental}
We also propose an alternative way for Gaussian kinematics that better fits the updated Lagrangian framework, which avoids the dependency on the total deformation gradient \( \bm F \). This approach also paves the way for physical material models that do not rely on employing \( \bm F \) as the strain measure.
Following conventions from computational fluid dynamics \cite{mckiver2003motion,chandrasekhar1967ellipsoidal}, the update rule for the world-space covariance matrix $\bm a$ can also be derived by discretizing the rate form of kinematics $\dot{\bm a}= (\nabla \bm v) \bm a + \bm a (\nabla \bm v)^T$: 
\begin{equation}
\bm a_p^{n+1} = \bm a_i^{n} + \Delta t (\nabla \bm  v_p \bm  a_p^n + \bm a_p^n \nabla  \bm  v_p^T).
\end{equation}
This formulation facilitates the incremental update of the Gaussian kernel shapes from time step $t^n$ to $t^{n+1}$ without the need to obtain \( \bm  F_p \). The rotation matrix $\bm R_p$ of each spherical harmonics basis can be incrementally updated in a similar manner. Starting from $\bm R^0_p = \bm I$, we extract the rotation matrix $\bm R_p^{n+1}$ from $(\bm I + \Delta t \bm v_p) \bm R_p^{n}$ using the polar decomposition.

\subsection{Internal Filling}
\label{sec:filling}
The internal structure is occluded by the object's surface, as the reconstructed Gaussians tend to distribute near the surface, resulting in inaccurate behaviors of volumetric objects. To fill particles into the void internal region, inspired by \citet{tang2023dreamgaussian}, we borrow the 3D opacity field from 3D Gaussians
\begin{equation}
\resizebox{.85\linewidth}{!}{
    $d(\bm x) = \sum_p \sigma_p \exp\left( -\frac{1}{2} (\bm x - \bm x_p)^T \bm A_p^{-1} (\bm x -\bm  x_p)\right)$.
    }
\end{equation}
\begin{wrapfigure}{r}{0.5\linewidth}
\vspace{-1.0em}
\hspace{-1.0em}
    \includegraphics[width=\linewidth]{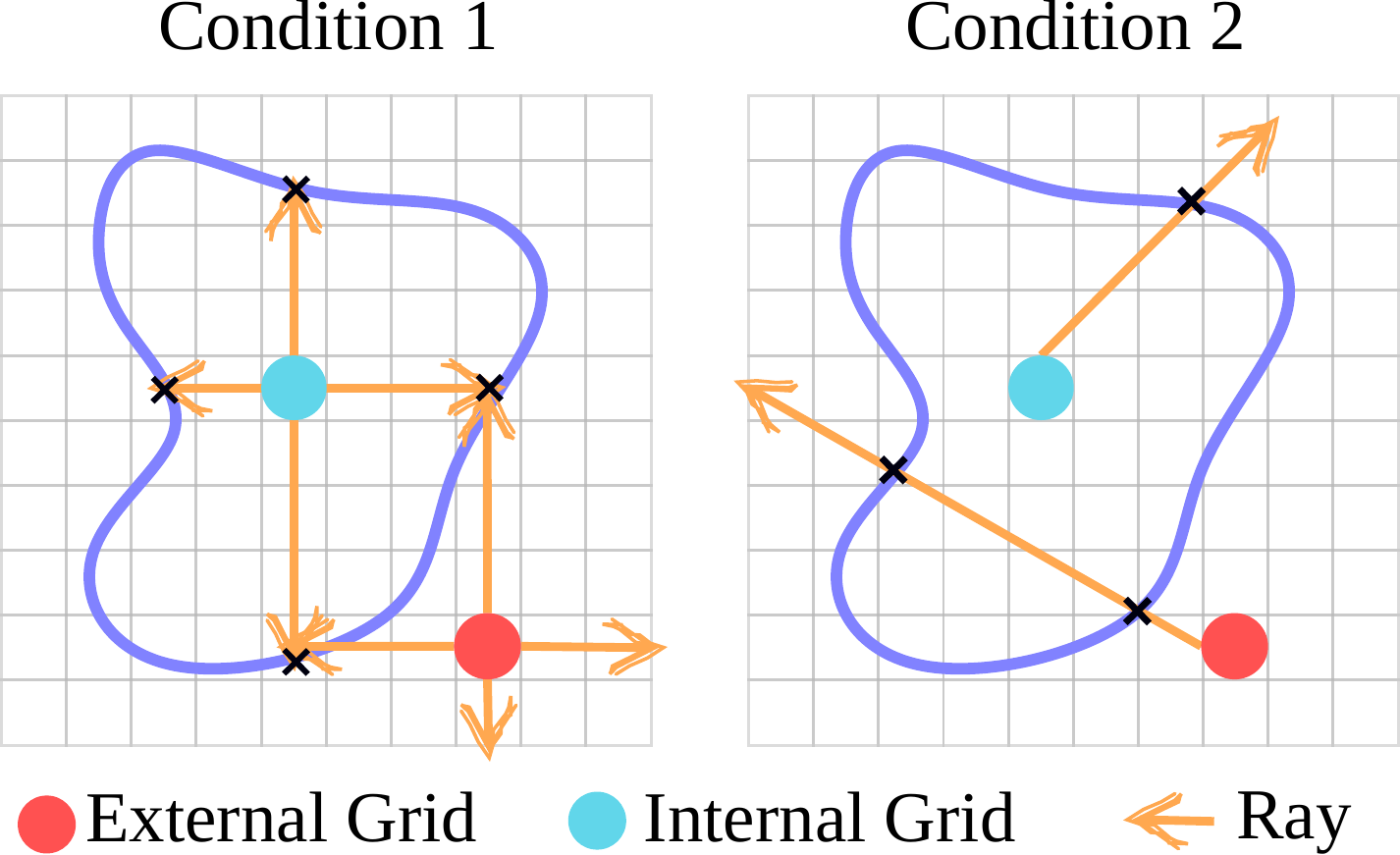}
    \label{fig: filling illustration}
\vspace{-0.5em}
\end{wrapfigure}
This continuous field is discretized onto a 3D grid. To achieve robust internal filling, we first define the concept of ``intersection'' within the opacity field, guided by a user-defined threshold $\sigma_{th}$. Specifically, we consider it an intersection when a ray passes from a lower opacity grid ($\sigma_i < \sigma_{th}$) to a higher opacity one ($\sigma_j > \sigma_{th}$). Based on this definition, we identify candidate grids by casting rays along 6 axes and checking intersections (condition 1). Rays originating from internal cells will always intersect with the surface. To further refine our selection of candidate grids, we employ an additional ray to assess the intersection number (condition 2), thus ensuring greater accuracy.

Visualization of these internal particles is also crucial as they may get exposed due to large deformation. Those filled particles inherit $\sigma_p, \mathcal{C}_p$ from their closet Gaussian kernels. Each particle's covariance matrix is initialized as $\operatorname{diag}(r^2_p, r^2_p, r^2_p)$, where $r$ is the particle radius calculated from its volume: $r_p = ({3V^0_p}/{4\pi})^{\frac{1}{3}}$. Alternatively, one may also consider employing generative models for internal filling, potentially leading to more realistic results. 

\subsection{Anisotropy Regularizer}
\label{sec:regularize}
The anisotropy of Gaussian kernels increases the efficiency of 3D representation while over-skinny kernels may point outward from the object surface under large deformations, leading to unexpected plush artifacts. We propose the following training loss during 3D Gaussian reconstruction:
\begin{equation}\label{eq:regularization}
    \mathcal{L}_{aniso} = \frac{1}{|\mathcal{P}|}\sum_{p\in \mathcal{P}}\max\{\max(\bm S_p) / \min(\bm S_p), r\} - r,
\end{equation}
where $\bm S_p$ are the scalings of 3D Gaussians \cite{kerbl20233d}. This loss essentially constrains that the ratio between the major axis length and minor axis length does not exceed $r.$ If desired, this term can be added to the training loss.

\section{Experiments}

In this section, we show the versatility of our approach across a wide range of materials. We also evaluate the effectiveness of our method across a comprehensive suite of benchmarks.

\subsection{Evaluation of Generative Dynamics}
\begin{figure*}
    \centering
    \includegraphics[width=\linewidth]{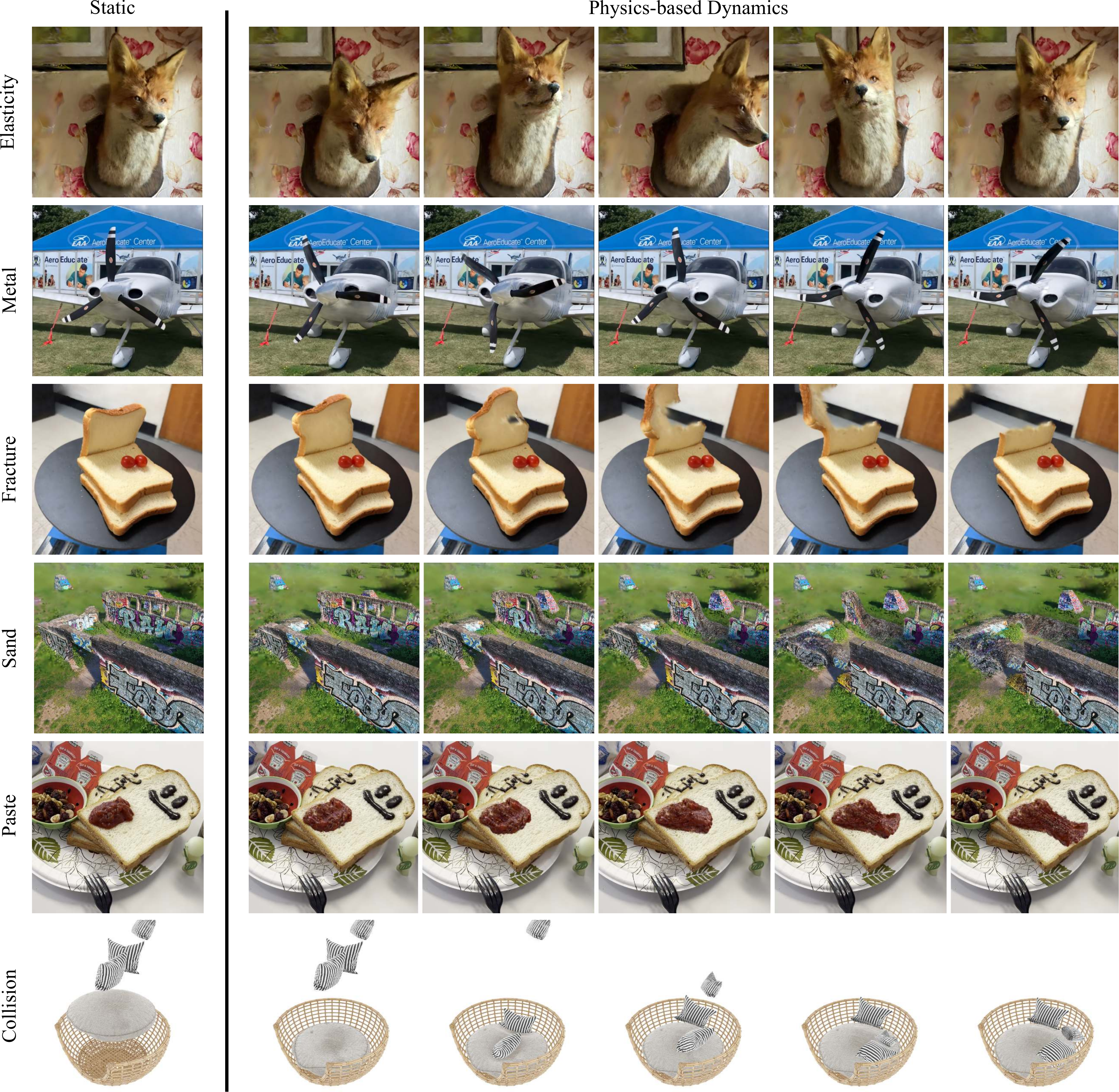}
    \caption{\textbf{Material Versatility.} We demonstrate exceptional versatility of our approach across a wide variety of examples: \textit{fox} (elastic entity), \textit{plane} (plastic metal), \textit{toast} (fracture), \textit{ruins} (granular material), \textit{jam} (viscoplastic material), and \textit{sofa suite} (collision).
    }
    \label{fig: various_material}
\end{figure*}
\paragraph{Datasets} We evaluate our method for generating diverse dynamics using several sources of input. In addition to the synthetic data (\emph{sofa suite}) generated by BlenderNeRF \cite{Raafat_BlenderNeRF_2023}, we utilize \emph{fox}, \emph{plane}, and \emph{ruins} from the datasets of Instant-NGP \cite{muller2022instant}, Nerfstudio \cite{tancik2023nerfstudio} and the DroneDeploy NeRF \cite{Pilkington2022dd}, respectively. 
Furthermore, we collect two real-world datasets (referred to as \emph{toast} and \emph{jam}) with an iPhone. Each scene contains 150 photos. The initial point clouds and camera parameters are obtained using COLMAP \cite{schoenberger2016mvs, schoenberger2016sfm}.

\vspace{-1em}
\paragraph{Simulation Setups}
We build upon the MPM from \citet{zong2023neural}.
To generate novel physics-based dynamics of a 3D Gaussian scene, we manually select a simulation region and normalize it to a cube with edge length 2. The internal particle filling can be performed before simulation. The cuboid simulation domain is discretized by a 3D dense grid. We selectively modify the velocities of specific particles to induce controlled movement. The remaining particles follow natural motion patterns governed by the established physical laws. 
All our experiments are performed on a 24-core 3.50GHz Intel i9-10920X machine with a Nvidia RTX 3090 GPU.

\vspace{-1em}
\paragraph{Results}
We simulate a wide range of physics-based dynamics. For each type of dynamics, we visualize one example with its initial scene and deformation sequence, as shown in \cref{fig: various_material}. Additional experiments are included in the supplemental document. The dynamics include: \textbf{Elasticity} refers to the property where the rest shape of the object remains invariant during deformation, representing the simplest form of daily-life dynamics. \textbf{Metal} can undergo permanent rest shape changes, which follows von-Mises plasticity model. \textbf{Fracture} is naturally supported by MPM simulation, where large deformations can cause particles to separate into multiple groups. \textbf{Sand} follows Druker-Prager plasticity model \cite{klar2016drucker}, which can capture granular-level frictional effects among particles.  \textbf{Paste} is modeled as viscoplastic non-Newtonian fluid, adhering to  Herschel-Bulkley plasticity model \cite{yue2015continuum}. \textbf{Collision} is another key feature of MPM simulation, which is automatically handled by grid time integration. Explicit MPM can be highly optimized to run on GPUs. We highlight that some of the cases can achieve real-time based on the $1/24$-s frame duration: \emph{plane} (30 FPS), \emph{toast} (25 FPS) and \emph{jam} (36 FPS). While utilizing FEM may further accelerate the elasticity simulation, it will involve an additional step of mesh extraction and lose the generalizability of MPM in inelasticity simulation.
% . Moreover, FEM is not a general simulator for inelasticity and requires additional effort for collision handling.

\begin{figure}
    \centering
    \includegraphics[width=\linewidth]{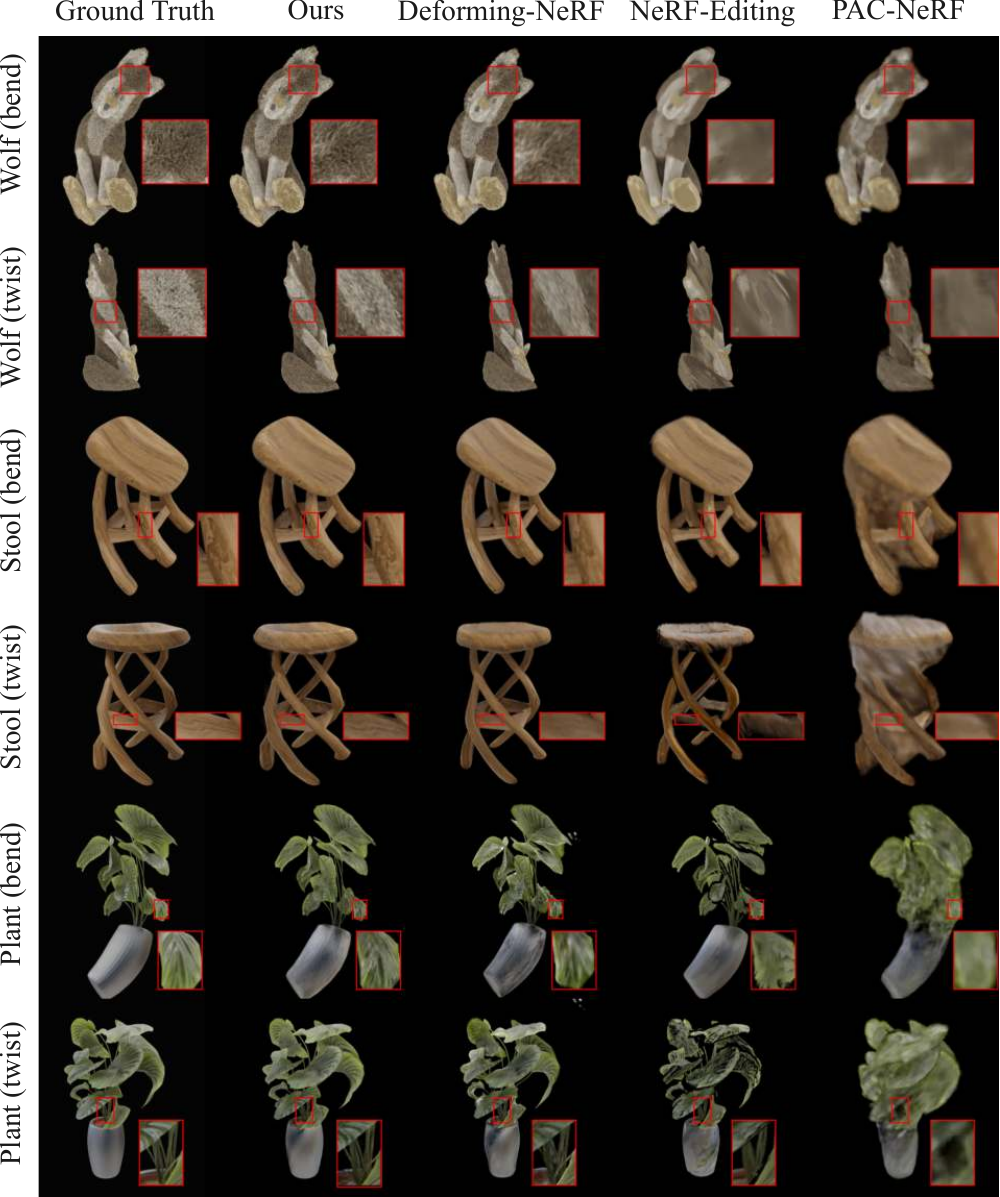}
    \caption{\textbf{Comparisons.} For each benchmark case, we select one test viewpoint and visualize all comparisons. We zoom in on some regions to highlight the ability of our method to maintain high-fidelity rendering quality after deformations. We use a black background to avoid the interference of the background.}
    \label{fig: comparison}
\end{figure}

\subsection{Lattice Deformation Benchmarks}

\paragraph{Dataset}
 Due to the absence of ground truth for post-deformation, we utilize BlenderNeRF \cite{Raafat_BlenderNeRF_2023} to synthesize several scenes, applying bending and twisting with the lattice deformation tool. For each scene, we create 100 multi-view renderings of the undeformed state for training, and 100 multi-view renderings of each deformed state to serve as ground truth for the deformed NeRFs. The lattice deformations are set as input to all methods for fair comparisons.

\vspace{-1em}
\paragraph{Comparisons}
We compare our method with several state-of-the-art NeRF frameworks that support manual deformations: \textbf{1) NeRF-Editing \cite{yuan2022nerf}} deforms NeRF using an extracted surface mesh, \textbf{2) Deforming-NeRF \cite{xu2022deforming}} utilizes a cage mesh for deformation, and \textbf{3) PAC-NeRF \cite{li2023pacnerf}} manipulates individual initial particles.

We show qualitative results in \cref{fig: comparison} and quantitative results in \cref{tab: comparison}. NeRF-Editing uses NeuS \cite{wang2021neus} as the scene representation, which is more suited for surface reconstructions rather than high-fidelity renderings. Consequently, its rendering quality is inherently lower than that of 3DGS. Furthermore, the deformation highly depends on the precision of the extracted surface mesh and the dilated cage mesh -- an overly tight mesh might not encompass the entire radiance field, while an excessively large one could result in a void border, as observed in the twisting stool and plant examples. 
Deforming-NeRF, on the other hand, provides clear renderings and potentially leads to enhanced results if higher-resolution deformation cages are provided. However, it employs a smooth interpolation from all cage vertices, thus filtering out fine local details and failing to match lattice deformations. PAC-NeRF is designed for simpler objects and textures in system identification tasks. While offering flexibility through its particle representation, it does not achieve high rendering fidelity. Our method utilizes both zero-order information (the deformation map) and first-order information (the deformation gradient) from each lattice cell. It outperforms the other methods across all cases, as high rendering qualities are well preserved after deformations. Although not primarily designed for editing tasks, this comparison showcases our method's significant potential for realistic editing of static NeRF scenes.
\begin{figure}
    \centering
    \includegraphics[width=\linewidth]{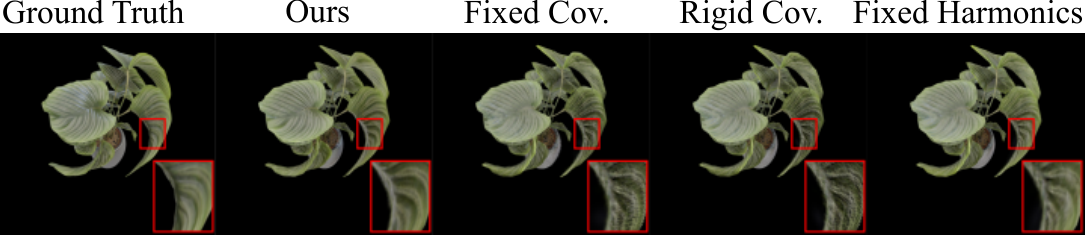}
    \caption{\textbf{Ablation Studies.} Non-extensible Gaussians can lead to severe visual artifacts during deformations. Although direct rendering the deformed Gaussian kernels can already obtain good results, additional rotations on spherical harmonics can improve the rendering quality.}
    \label{fig:deformation-ablation}
\end{figure}

\begin{table}[t]
\caption{We synthesize a lattice deformation benchmark dataset to compare with baselines and conduct ablation studies to validate our design choices. PSNR scores are reported (higher is better). Our method outperforms the others across all cases.}\label{tab: comparison}
\resizebox{\linewidth}{!}{
\centering
\begin{tabular}{lllllll}
\toprule
\multicolumn{1}{l}{Test Case}                                               & \multicolumn{2}{c}{Wolf} & \multicolumn{2}{c}{Stool} & \multicolumn{2}{c}{Plant} \\
\multicolumn{1}{l}{Deformation Operator}                                               & Bend       & Twist       & Bend        & Twist       & Bend        & Twist       \\
\midrule
NeRF-Editing \cite{yuan2022nerf}   &     26.74  &  24.37   &    25.00  &    21.10 &  19.85 &   19.08  \\
Deforming-NeRF \cite{xu2022deforming}   &   21.65  &  21.72    &   22.32  &    21.16  &   17.90  &     18.63 \\
PAC-NeRF  \cite{li2023pacnerf}       &  26.91   &  25.27     &  21.83    &   21.26  &  18.50    &    17.78  \\
\midrule
Fixed Covariance    &  26.77   &  26.02 &   29.94    &   25.31  &  23.95   &     23.09 \\
Rigid Covariance    &    26.84    &  26.16 & 30.28     &   25.70   &    24.09  &   23.53  \\
Fixed Harmonics   &    26.83  & 26.02 & 30.87    &    25.75   &     25.09   &  23.69 \\
Ours & \textbf{26.96} & \textbf{26.46}     &   \textbf{31.15}    &  \textbf{26.15}    &  \textbf{25.81}   & \textbf{23.87} \\
\bottomrule
\end{tabular}
}
\end{table}

\vspace{-1em}
\paragraph{Ablation Studies}
We further conduct several ablation studies on these benchmark scenes to validate the necessity of the kinematics of Gaussian kernels and spherical harmonics: \textbf{1) Fixed Covariance} only translates the Gaussian kernels. \textbf{2) Rigid Covariance} only applies rigid transformations on the Gaussians, as assumed in \citet{luiten2023dynamic}. \textbf{3) Fixed Harmonics} does not rotate the orientations of spherical harmonics, as assumed in \citet{wu20234d}.

Here we visualize one example in \cref{fig:deformation-ablation}. We can observe that Gaussians will not properly cover the surface after deformation if they are non-extensible, leading to severe visual artifacts. Enabling the rotation of spherical harmonics can slightly improve the consistency with the ground truth. We include quantitative results on all test cases in \cref{tab: comparison}, which shows that all these enhancements are needed to achieve the best performance of our method.

\subsection{Additional Qualitative Studies}

\paragraph{Internal Filling}
\begin{figure}[t]
    \centering
    \includegraphics[width=\linewidth]{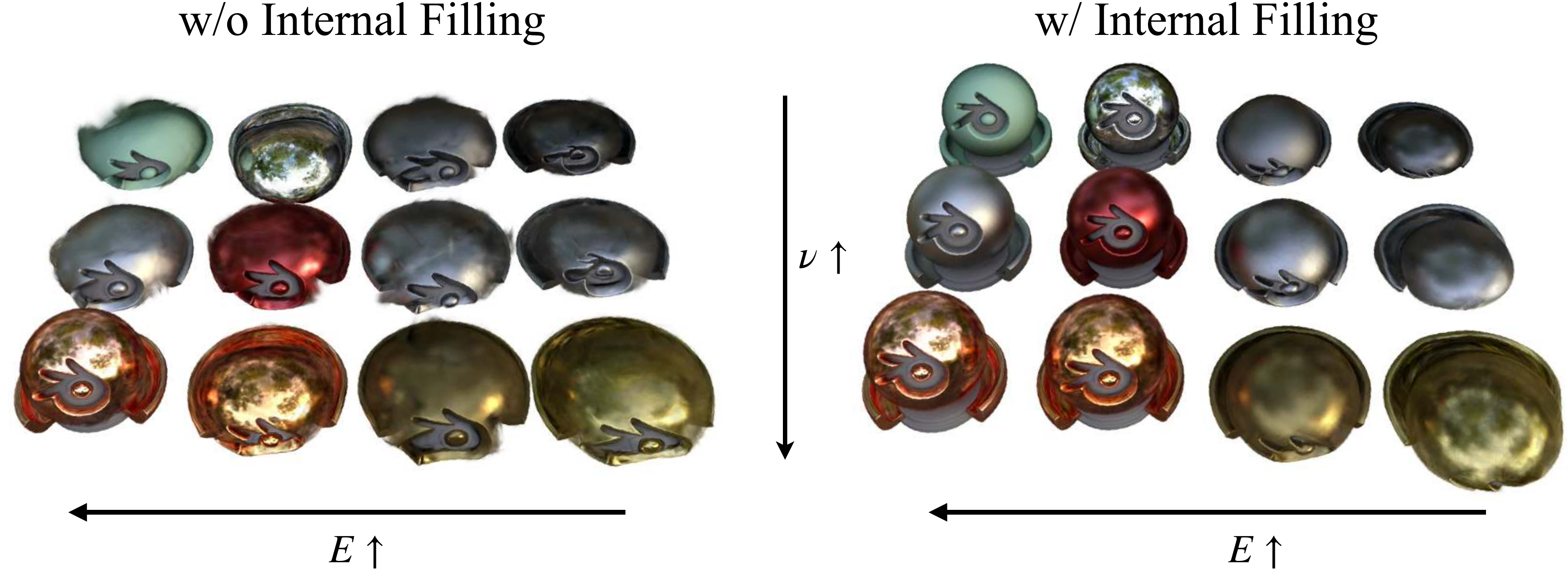}
    \caption{\textbf{Internal filling} enables more realistic simulation results. Our method also supports flexible control of dynamics via material parameters. A larger Young's modulus $E$ indicates higher stiffness while a larger poission ratio $\nu$ leads to better volume preservation.}
    \label{fig: filling_comparison}
\end{figure}
Typically, the 3D Gaussian splatting framework focuses on the surface appearance of objects and often fails to capture their internal structure. Consequently, the interior of the modeled object remains hollow, resembling a mere shell. This is usually not true in the real world, leading to unrealistic simulation results. To address this challenge, we introduce an internal filling method leveraging a reconstructed density field, which is derived from the opacity of Gaussian kernels. 
\cref{fig: filling_comparison} showcases our simulation results with varying physical parameters. Objects devoid of internal particles tend to collapse when subjected to gravity forces, irrespective of the material parameters used. In contrast, our approach assisted by internal filling allows for nuanced control over object dynamics, effectively adjusting to different material characteristics. 

\vspace{-1em}
\paragraph{Volume Conservation}
\begin{figure}[t]
    \centering
    \includegraphics[width=\linewidth]{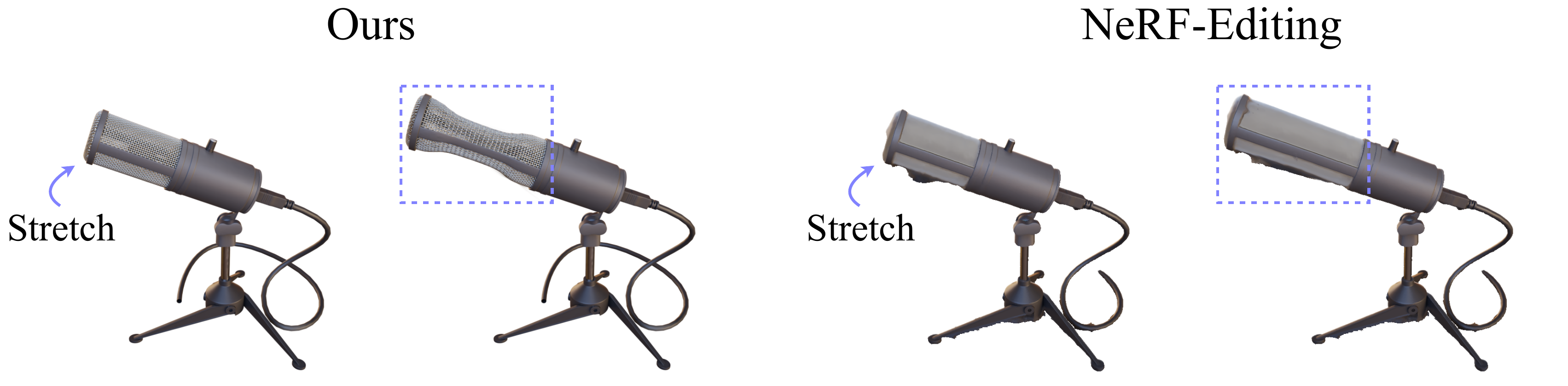}
    \caption{\textbf{Volume Conservation.} Compared to the geometry-based editing method \cite{yuan2022nerf}, our physics-based method is able to capture volumetric behaviors, leading to more realistic dynamics.}
    \label{fig: volume_conservation}
\end{figure}

\begin{figure}[t]
    \centering
    \includegraphics[width=\linewidth]{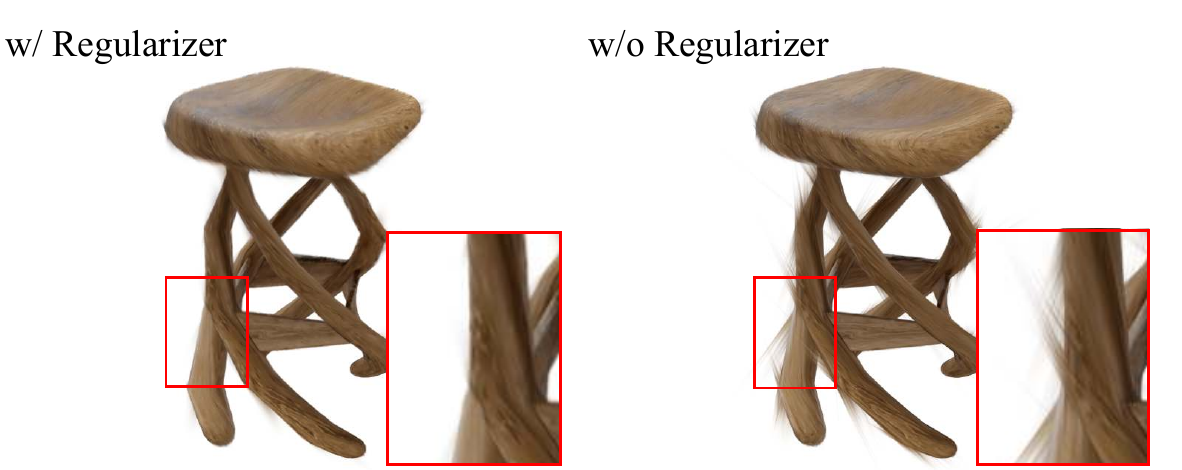}
    \caption{\textbf{Anisotropy Regularizer.} We introduce an anisotropy constraint for Gaussian kernels, effectively enhancing the fidelity of the Gaussian-based representation under dynamic conditions.}
    \label{fig: regularization}
\end{figure}

Existing approaches to NeRF manipulation focus primarily on geometric adjustments without incorporating physical properties. A key attribute of real-world objects is their inherent ability to conserve volume during deformation. In \cref{fig: volume_conservation}, we conduct a comparison study between our method and NeRF-Editing \cite{yuan2022nerf}, which utilizes surface As-Rigid-As-Possible (ARAP) deformation \cite{sorkine2007rigid}. Unlike NeRF-Editing, our approach accurately captures and maintains the volume of the deformed objects.
\vspace{-1em}
\paragraph{Anisotropy Regularizer}
3D Gaussian models inherently represent anisotropic ellipsoids. However, excessively slender Gaussian kernels can lead to burr-like visual artifacts, especially pronounced during large deformations To tackle this issue, we introduce an additional regularization loss \cref{eq:regularization} to constrain anisotropy. As demonstrated in \cref{fig: regularization}, this additional loss function effectively mitigates the artifacts induced by extreme anisotropy.

% \subsection{Various Materials}

\section{Discussion}

\paragraph{Conclusion} This paper introduces \METHOD, a unified simulation-rendering pipeline that generates physics-based dynamics and photo-realistic renderings simultaneously and seamlessly. 

\vspace{-1em}
\paragraph{Limitation} In our framework, the evolution of shadows is not considered, and material parameters are manually set. Automatic parameter assignment could be derived from videos by combining GS segmentation \cite{cen2023segment, ye2023gaussian} with a differentiable MPM simulator. Additionally, incorporating geometry-aware 3DGS reconstruction methods \cite{guedon2023sugar} could enhance generative dynamics.
Future work will also explore handling more versatile materials like liquids and integrating more intuitive user controls, possibly leveraging advancements in Large Language Models (LLMs).

\vspace{-1em}
\paragraph{Acknowledgements} We thank Ying Nian Wu and Feng Gao for useful discussions. We acknowledge support from NSF (2301040, 2008915, 2244651, 2008564, 2153851, 2023780), UC-MRPI, Sony, Amazon, and TRI.

\newpage
{
    \small
    \bibliographystyle{ieeenat_fullname}
    \bibliography{main}
}

\newpage
\appendix
\section*{\Large Appendix}
\input{supplemental}

\end{document}

%% file: supplemental.tex
\section{MPM Algorithm}
In MPM, a continuum body is discretized into a set of Lagrangian particles $p,$ and time is discretized into a sequence of time steps $t = 0, t^1, t^2, ...$. Here we take a fixed stepsize $\Delta t,$ so $t^n = n \Delta t.$ 

At each time step, masses and momentums on particles are first transferred to grid nodes. Grid velocities are then updated using forward Euler's method and transferred back to particles for subsequent advection. Let $m_p,$ $\boldsymbol{x}_p^n$, $\boldsymbol{v}_p^n$, $\boldsymbol{F}_p^n$, $\boldsymbol{\tau}_p^n$, and $\boldsymbol{C}_p^n$ denote the mass, position, velocity, deformation gradient, Kirchhoff stress, and affine momentum on particle $p$ at time $t_n$. Let $m_i$, $\boldsymbol{x}_i^n$ and $\boldsymbol{v}_i^n$ denote the mass, position, and velocity on grid node $i$ at time $t^n$. Here, particle masses are invariant due to mass conservation law. Let $m_i^n$, $\boldsymbol{x}_i^n$ and $\boldsymbol{v}_i^n$ denote the mass, position, and velocity on grid node $i$ at time $t^n$. We summarize the explicit MPM algorithm as follows: 

\begin{enumerate}
    \item \textbf{Transfer Particles to Grid.} Transfer mass and momentum from particles to grids as 
    \begin{equation}
    \begin{aligned}
        m_{{i}}^n &=\sum_p w_{{i} p}^n m_p, \\
        m_i^n\boldsymbol{v}_{{i}}^n &= \sum_p w_{{i} p}^n m_p\left(\boldsymbol{v}_p^n+\boldsymbol{C}_p^n\left(\boldsymbol{x}_{{i}}-\boldsymbol{x}_p^n\right)\right).
    \end{aligned}
    \end{equation}
    We adopt the APIC scheme \cite{jiang2015affine} for momentum transfer.
    \item \textbf{Grid Update.} Update grid velocities based on forces at the next timestep by 
    \begin{equation}
        \boldsymbol{v}_i^{n+1} = \boldsymbol{v}_i^{n} - \frac{\Delta t}{m_i} \sum_p \boldsymbol{\tau}_p^n \nabla w_{i p}^n V_p^0 + \Delta t \boldsymbol{g}.
    \end{equation}
    \item \textbf{Transfer Grid to Particles.} Transfer velocities back to particles and update particle states. 
    \begin{equation}
    \begin{aligned}
        \boldsymbol{v}_p^{n+1}&=\sum_i \boldsymbol{v}_{{i}}^{n+1} w_{{i} p}^n, \\
        \boldsymbol{x}_p^{n+1} &= \boldsymbol{x}_p^{n} + \Delta t \boldsymbol{v}_p^{n+1}, \\
        {\boldsymbol{C}}_p^{n+1}&=\frac{12}{\Delta x^2(b+1)} \sum_{{i}} w_{{i} p}^n \boldsymbol{v}_i^{n+1} \left(\boldsymbol{x}_{{i}}^n-\boldsymbol{x}_p^n\right)^T,\\
        \nabla \bm v_p^{n+1} &= \sum_i \boldsymbol{v}_{{i}}^{n+1} {\nabla w_{{i} p}^n}^{T},\\
        \boldsymbol{F}_p^{\text{E, tr}} &= (\bm I + \nabla \bm v_p^{n+1}) \bm F^{E, n},\\
        \boldsymbol{F}_p^{E,n+1} &= \mathcal{Z}(\boldsymbol{F}_p^{\text{E, tr}}), \\
        \boldsymbol{\tau}_p^{n+1} &= \boldsymbol{\tau}(\boldsymbol{F}_p^{E,n+1}).
    \end{aligned}
    \end{equation}
    % $\boldsymbol{v}_p^{n+1}=\sum \boldsymbol{v}_{{i}}^{n+1} w_{{i} p}^n,$ $\boldsymbol{x}_p^{n+1} = \boldsymbol{x}_p^{n} + \Delta t \boldsymbol{v}_p^{n+1},$ $${\boldsymbol{C}}_p^{n+1}=\frac{12}{\Delta x^2(b+1)} \sum_{{i}} w_{{i} p}^n \boldsymbol{v}_i^{n+1} \left(\boldsymbol{x}_{{i}}^n-\boldsymbol{x}_p^n\right)^T,$$
    % $$\boldsymbol{F}_p^{\text{trial}, n+1} =\left(\mathbf{I}+\Delta t \sum_i \boldsymbol{v}_i^{n+1}\left(\nabla w_{i p}^n\right)^T\right) \boldsymbol{F}_p^{E,n},$$
    % $$\boldsymbol{F}_p^{E,n+1} = \text{returnMap}(\boldsymbol{F}_p^{\text{trial}, n+1}) \text{ and } \boldsymbol{\tau}_p^{n+1} = \boldsymbol{\tau}(\boldsymbol{F}_p^{E,n+1}).$$ 
    Here $b$ is the B-spline degree, and $\Delta x$ is the Eulerian grid spacing. The computation of the return map $\mathcal{Z}$ and the Kirchhoff stress $\boldsymbol{\tau}$ is outlined in \cref{sec:constitutive-model}.  We refer the readers to \cite{jiang2016material} for the detailed derivations from the continuous conservation law to its MPM discretization.
\end{enumerate}

\section{Elasticity and Plasticity Models}
\label{sec:constitutive-model}
We adopt the constitutive models used in \cite{zong2023neural}.
We list the models used for each scene in \cref{tab: model_settings} and summarize all the parameters needed in discussing the constitutive models in \cref{tab: material_parameters}.
\begin{table}[t]
\centering
\caption{\textbf{Model Settings.}} \label{tab: model_settings}
\begin{tabular}{lll}
\hline
Scene             & Figure          & Constitutive Model \\ \hline
Vasedeck          & \cref{fig:teaser}          & Fixed corotated    \\
Ficus             & \cref{fig:overview}          & Fixed corotated    \\
Fox               & \cref{fig: various_material}          & Fixed corotated    \\
Plane             & \cref{fig: various_material}          & von Mises          \\
Toast             & \cref{fig: various_material}          & Fixed corotated    \\
Ruins             & \cref{fig: various_material}          & Drucker-Prager     \\
Jam               & \cref{fig: various_material}          & Herschel-Bulkley   \\
Sofa Suite        & \cref{fig: various_material}          & Fixed corotated    \\
Materials         & \cref{fig: filling_comparison}          & Fixed corotated    \\
Microphone        & \cref{fig: volume_conservation}          & Neo-Hookean        \\
Bread             & \cref{fig:additional_dynamics} & Fixed corotated    \\
Cake              & \cref{fig:additional_dynamics} & Herschel-Bulkley   \\
Can               & \cref{fig:additional_dynamics} & von Mises          \\
Wolf              & \cref{fig:additional_dynamics} & Drucker-Prager   \\
\hline
\end{tabular}
\end{table}

\begin{table}[t]
\centering
\caption{\textbf{Material Parameters.}} \label{tab: material_parameters}
\begin{tabular}{lll}
\hline Notation& Meaning & Relation to $E, \nu$ \\
\hline 
$E$         & Young's modulus   & / \\
$\nu$       & Poisson's ratio   & / \\
$\mu$       & Shear modulus     &  $\mu=\frac{E}{2(1+\nu)}$ \\
$\lambda$   & Lamé modulus      & $\lambda=\frac{E \nu}{(1+\nu)(1-2 \nu)} $\\
$\kappa$    & Bulk modulus      & $\kappa = \frac{E}{3(1-2 \nu)}$\\
\hline
\end{tabular}
\end{table}

\begin{figure*}
    \centering
    \includegraphics[width=\linewidth]{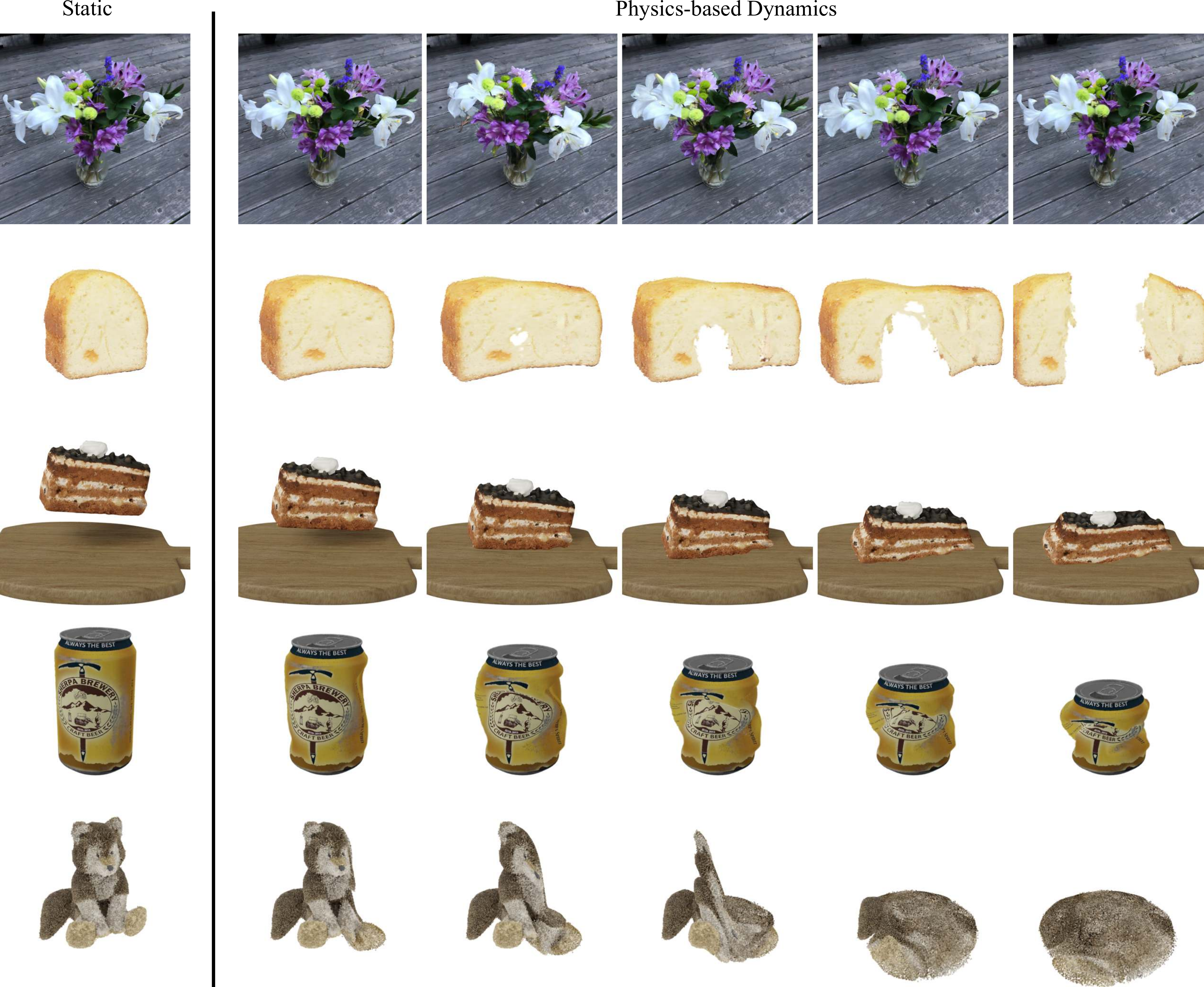}
    \caption{\textbf{Additional Evaluation.} Examples from top to bottom are: \textit{vasedeck} (elastic entity), \textit{bread} (fracture), \textit{cake} (viscoplastic material), \textit{can} (metal) and \textit{wolf} (granular material).}
    \label{fig:additional_dynamics}
\end{figure*}

In all plasticity models used in our work, the deformation gradient is multiplicatively decomposed into $\boldsymbol{F}=\boldsymbol{F}^E \boldsymbol{F}^P$ following some yield stress condition. A hyperelastic constitutive model is applied to $\boldsymbol{F}^E$ to compute the Kirchhoff stress $\boldsymbol{\tau}.$ For a pure elastic continuum, we simply take $\boldsymbol{F}^E=\boldsymbol{F}.$

\subsection{Fixed Corotated Elasticity}
The Kirchhoff stress $\boldsymbol{\tau}$ is defined as 
\begin{equation}
\boldsymbol{\tau} = 2  \mu(\boldsymbol{F}^E-\boldsymbol{R}) {\boldsymbol{F}^E}^{T}+\lambda(J-1) J,
\end{equation}
where $\boldsymbol{R} = \boldsymbol{U} \boldsymbol{V}^T$ and $\boldsymbol{F}^E = \boldsymbol{U}\boldsymbol{\Sigma}\boldsymbol{V}^T$ is the singular value decomposition of elastic deformation gradient. $J$ is the determinant of $\boldsymbol{F}^E$ \cite{jiang2015affine}.

\subsection{StVK Elasticity}
The Kirchhoff stress $\boldsymbol{\tau}$ is defined as 
\begin{equation}
    \boldsymbol{\tau}=\boldsymbol{U} \left(2 \mu \boldsymbol{\epsilon}+\lambda \operatorname{sum}(\boldsymbol{\epsilon}) \mathbf{1}\right) \boldsymbol{V}^T,
\end{equation}
where $\boldsymbol{\epsilon}=\log (\boldsymbol{\Sigma})$ and $\boldsymbol{F}^E = \boldsymbol{U}\boldsymbol{\Sigma}\boldsymbol{V}^T$ \cite{klar2016drucker}.

\subsection{Neo-Hookean Elasticity}
The Kirchhoff stress $\boldsymbol{\tau}$ is defined as 
\begin{equation}
\boldsymbol{\tau} = \mu(\boldsymbol{F}^E {\boldsymbol{F}^E}^T - \boldsymbol{I}) + \log(J) \boldsymbol{I},
\end{equation}
where $J$ is the determinant of $\boldsymbol{F}^E$ \cite{jiang2015affine}.

\subsection{Drucker-Prager Plasticity}
The return mapping of Drucker-Prager plasticity for sand \cite{klar2016drucker} is, given $\boldsymbol{F} = \boldsymbol{U}\boldsymbol{\Sigma}\boldsymbol{V}^T$ and $\boldsymbol{\epsilon}=\log (\boldsymbol{\Sigma}),$ 
\begin{equation}
    \boldsymbol{F}^E = \boldsymbol{U} \mathcal{Z}(\boldsymbol{\Sigma}) \boldsymbol{V}^T,
\end{equation}
% $$\boldsymbol{F}^E = \boldsymbol{U} \mathcal{Z}(\boldsymbol{\Sigma}) \boldsymbol{V}^T.$$
% $$
\begin{equation}
\mathcal{Z}(\boldsymbol{\Sigma})=\left\{\begin{array}{ll}
\mathbf{1}, & \operatorname{sum}(\boldsymbol{\epsilon})>0, \\
\boldsymbol{\Sigma}, & \delta \gamma \leq 0, \text { and } \operatorname{sum}(\boldsymbol{\epsilon}) \leq 0, \\
\exp \left(\boldsymbol{\epsilon}-\delta \gamma \frac{\hat{\epsilon}}{\|\hat{\epsilon}\|}\right), & \text { otherwise. }
\end{array}\right.
\end{equation}
% $$
Here $\delta \gamma=\|\hat{\boldsymbol{\epsilon}}\|+\alpha \frac{(d \lambda+2  \mu) \operatorname{sum}(\boldsymbol{\epsilon})}{2  \mu},$ $\alpha=\sqrt{\frac{2}{3}} \frac{2 \sin \phi_f}{3-\sin \phi_f},$ and $\phi_f$ is the friction angle. $\hat{\epsilon} = \operatorname{dev}(\epsilon).$
\subsection{von Mises Plasticity}
Similar to Drucker-Prager plasticity, given $\boldsymbol{F} = \boldsymbol{U}\boldsymbol{\Sigma}\boldsymbol{V}^T$ and $\boldsymbol{\epsilon}=\log (\boldsymbol{\Sigma}),$ 
\begin{equation*}
    \boldsymbol{F}^E = \boldsymbol{U} \mathcal{Z}(\boldsymbol{\Sigma}) \boldsymbol{V}^T,
\end{equation*}
where
\begin{equation}
\mathcal{Z}(\boldsymbol{\Sigma})=\left\{\begin{array}{ll}
\boldsymbol{\Sigma}, & \delta \gamma \leq 0, \\
\exp \left(\boldsymbol{\epsilon}-\delta \gamma \frac{\hat{\epsilon}}{\|\boldsymbol{\epsilon}\|}\right), & \text { otherwise, }
\end{array}\right.
\end{equation}
and $\delta \gamma=\|\hat{\boldsymbol{\epsilon}}\|_F-\frac{\tau_Y}{2  \mu}.$ Here $\tau_Y$ is the yield stress. 

\subsection{Herschel-Bulkley Plasticity}
We follow \citet{yue2015continuum} and take the simple case where $h=1.$ Denote $\boldsymbol{s}^{\text{trial}} = \operatorname{dev}(\boldsymbol{\tau}^{\text{trial}}),$ and $s^{\text{trial}} = ||\boldsymbol{s}^{\text{trial}}||.$ The yield condition is $\Phi(s)=s-\sqrt{\frac{2}{3}} \sigma_Y \leq 0.$ If it is violated, we modify $s^{\text{trial}}$ by $$
s=s^{\text{trial}}-\left(s^{\text{trial}}-\sqrt{\frac{2}{3}} \sigma_Y\right) /\left(1+\frac{\eta}{2  \mu \Delta t}\right).
$$
$\boldsymbol{s}$ can then be recovered as $\boldsymbol{s} = s \cdot \frac{\boldsymbol{s}^{\text{trial}}}{||\boldsymbol{s}^{\text{trial}}||}.$
Define $\boldsymbol{b}^E = \boldsymbol{F}^E {\boldsymbol{F}^E}^T.$ The Kirchhoff stress $\boldsymbol{\tau}$ is computed as 
$$\boldsymbol{\tau} =\frac{\kappa}{2}\left(J^2-1\right) \boldsymbol{I}+ \mu \operatorname{dev}\left[\operatorname{det}(\boldsymbol{b}^E)^{-\frac{1}{3}} \boldsymbol{b}^E\right].$$

\section{Additional Evaluations}
We present additional evaluations of our method in \cref{fig:additional_dynamics}. The \textit{vasedeck} data is from the Nerf dataset \cite{mildenhall2021nerf} and the others are synthetic data, generated using BlenderNeRF \cite{Raafat_BlenderNeRF_2023}.